\documentclass[]{aa}  

\usepackage{natbib}
\bibpunct{(}{)}{;}{a}{}{,}
\usepackage{graphicx}
\usepackage{revsymb}	
\usepackage{amsmath}	
\usepackage[usenames]{color}	
\usepackage{epstopdf}	
\usepackage{txfonts}
\usepackage{amssymb}
\usepackage{color}
\usepackage[justification=centering]{caption}
\usepackage{geometry} 
\usepackage[parfill]{parskip} 
\usepackage{amssymb}
\usepackage{slantsc}
\usepackage{array}
\usepackage{url}
\usepackage{amsfonts}
\usepackage[colorlinks=true,linkcolor=black, urlcolor=black, citecolor=black]{hyperref}
\usepackage{hyperref} 
\usepackage{sidecap}
\usepackage{graphicx}
\usepackage{xcolor}
\usepackage{nicefrac}
\usepackage{subfigure}
\usepackage{epsfig}
\colorlet{rouge}{red!70!darkgray}

\begin{document}
\title{Revisiting Kepler-444}
\subtitle{Part I. Seismic modelling and inversions of stellar structure}
\author{G. Buldgen\inst{1} \and M. Farnir\inst{2} \and C. Pezzotti\inst{1} \and P. Eggenberger\inst{1} \and S. J. A. J. Salmon\inst{2} \and J. Montalban\inst{3} \and J. W. Ferguson\inst{4} \and S. Khan \inst{3} \and V. Bourrier\inst{1} \and B. M. Rendle\inst{3} \and G. Meynet\inst{1}\and A. Miglio\inst{3} \and A. Noels\inst{2}}
\institute{Observatoire de Genève, Université de Genève, 51 Ch. Des Maillettes, CH$-$1290 Sauverny, Suisse \and STAR Institute, Université de Liège, Allée du Six Août 19C, B$-$4000 Liège, Belgium \and School of Physics and Astronomy, University of Birmingham, Edgbaston, Birmingham B15 2TT, UK \and Department of Physics, Wichita State University, Wichita, KS 67260-0032.}
\date{April, 2019}
\abstract{The CoRoT and \textit{Kepler} missions have paved the way for synergies between exoplanetology and asteroseismology. The use of seismic data helps providing stringent constraints on the stellar properties which directly impact the results of planetary studies. Amongst the most interesting planetary systems discovered by \textit{Kepler}, Kepler-444 is unique by the quality of its seismic and classical stellar constraints. Its magnitude, age and the presence of $5$ small-sized planets orbiting this target makes it an exceptional testbed for exoplanetology.}
{We aim at providing a detailed characterization of Kepler-444, focusing on the dependency of the results on variations of key ingredients of the theoretical stellar models. This thorough study will serve as a basis for future investigations of the planetary evolution of the system orbiting Kepler-444.}
{We use local and global minimization techniques to study the internal structure of the exoplanet-host star Kepler-444. We combine seismic observations from the \textit{Kepler} mission, \textit{Gaia} DR$2$ data and revised spectroscopic parameters to precisely constrain its internal structure and evolution.}{We provide updated robust and precise determinations of the fundamental parameters of Kepler-444 and demonstrate that this low-mass star bore a convective core during a significant portion of its life on the main-sequence. Using seismic data, we are able to estimate the lifetime of the convective core to approximately $8$ Gyr out of the $11$ Gyr of the evolution of Kepler-444. The revised stellar parameters found by our thorough study are $\rm{M}=0.754\pm 0.03\rm{M}_{\odot}$, $\rm{R}=0.753\pm 0.01\rm{R}_{\odot}$, $\rm{Age}=11\pm 1$ Gy.}{}
\keywords{Stars: interiors - Stars: oscillations - Stars: fundamental parameters - Asteroseismology - Planetary systems}
\maketitle
\section{Introduction}

With the advent of space-based photometry missions such as CoRoT \citep{Baglin}, \textit{Kepler} \citep{Borucki} and TESS \citep{Ricker2015}, and with the future launch of the PLATO mission \citep{Rauer}, asteroseismology has been established as a standard discipline to derive fundamental stellar properties. These capabilities led naturally to synergies between stellar seismologists and exoplanetologists, as planetary detections almost exclusively use stellar-dependent approaches. Consequently, a better characterization of the host star also significantly improves the accuracy and precision of the derived planetary parameters. 

Good examples of such synergies are found in \citet{JCDKOI2010,Batalha2011,Huber2013,Huber2013Science,Huber2018,Campante2018}, where the use of stellar seismology can be seen as a key component of the studies of the planetary system. Moreover, the importance of stellar evolution for planets is not only limited to the determination of stellar fundamental parameters, but also to the dynamical evolution of planetary systems and the potential evaporation of planetary atmospheres. In this context, understanding transport properties of both angular momentum and chemical elements is crucial, as transport will affect rotation, which will affect activity, which in turn will alter the planetary properties. 

In this study, we revisit one of the most well-known planet-host stars for which a detailed seismic characterization has been carried out by \citet{Campante2015}, Kepler-444 (also identified as HIP 94931, KIC 6278762, KOI-3158, and LHS 3450). Kepler-444 has been extensively studied in recent years, with attempts to put upper limits on the planetary masses and densities to constrain the formation history of the system \citep{Dupuy2016,Mills2017,Hadden2017}. It is composed of $5$ small planets, orbiting within $0.1$ AU of their host star with periods lower than $10$ days. Their estimated densities imply the presence of volatile elements and the fact that they transit a luminous star ($\rm{M}_{G}=8.64$, with $\rm{M}_{G}$ the mean photometric magnitude of \textit{Gaia} in G-band as described in \citet{Evans2018}) make them excellent targets for in-depth atmospheric characterization. Such tightly packed planetary systems are particularly interesting in terms of formation history, as some studies suggest that such systems should form locally while the low density of some of these planets, especially in the case of Kepler-444 argues for a formation of the two outermost planets behind the snowline followed by a migration phase \citep{Mills2017}. Kepler-444 is however rather specific as the star is actually a triple system with a pair of spatially unresolved M-dwarf companions \citep{Campante2015, Dupuy2016} which could have influenced the protoplanetary disk \citep{Kraus2016,Hirsch2017,Bazso2017}. From a planetary perspective,  Kepler-444 is not unique, as many other of such systems have already been observed and studied to characterize the correlation between low stellar metallicity and the occurrence of multi-planet systems \citep{Hobson2017, Brewer2018}. These systems of packed, small, inner-planets are also of particular interest as they may exhibit tidally-stressed plate tectonics \citep{Zanazzi2019}.

In addition, Kepler-444 is of particular interest to study the atmospheric properties of exoplanets close to their host stars. Recently, \citet{Bourrier2017} detected a strong variation of the HI Ly$-\alpha$ signal during transiting and non-transiting periods of the system. These strong variations could either be due to stellar variability or to the presence of a strong hydrogen exosphere around the two outermost exoplanets. The hypothesis of stellar activity, although supported by the presence of variations during a non-transiting period, is still unlikely given the old age of the star. However, an increased stellar activity would of course also have dramatic consequences for the evaporation of the planetary atmospheres and could be of importance when studying other compact systems. Meanwhile, the confirmation of the planetary origin of the signal would imply that Kepler-444e and Kepler-444f could be the first known ocean planets with vast amounts of water at their surface to be able to replenish a hydrogen rich exosphere capable of reproducing the strong HI Ly$-\alpha$ variations \citep{Bourrier2017}. 

In this first study, we rather focus on the details of the stellar seismic modelling, e.g. the variations of the transport of chemicals, opacity and abundance tables. We use the recently published \textit{Gaia} parallaxes \citep{GAIA2016,GAIA2018} as well as revised spectroscopic parameters of Kepler-444 and discuss their impact on the stellar fundamental parameters. We carry out seismic inversions of the stellar mean density following \citet{ReeseDens}, \cite{Buldgen} and \cite{Buldgen2018}. Section \ref{SecForward} lays out the seismic and classical constraints used in the forward modelling, the minimization strategy as well as the model properties and the various physical ingredients that have been tested in our study. Section \ref{SecStrucInv} presents the results of mean density inversions applied to Kepler-444 and discusses the model-dependency and the impact of various surface corrections and Section \ref{SecConvCore} discusses the details of the core conditions of the star, namely the survival of a convective core during an extended portion of the main-sequence. Section \ref{SecConc} summarizes our results and discusses future studies of Kepler-444 and their potential impact in the context of the TESS \citep{Ricker2015} and PLATO missions \citep{Rauer}.

Our future studies will use these results to provide a modelling of Kepler-444 including transport processes of angular momentum following the prescriptions for magnetic instabilities used by \citet{Eggenberger2010}. These models will then be used to constrain the dynamical evolution of the planetary system, using the approach \citet{Privitera2016I,Privitera2016AII,Privitera2016III,Meynet2017} and including the effects of dynamical tides following \citet{Rao2018}.

\section{Forward modelling}\label{SecForward}

In this section, we discuss the forward modelling procedures used to study the internal structure of Kepler-444 and determine its fundamental parameters. We used the Liège stellar evolution code \citep[CLES;][]{ScuflaireCles} and the Liège stellar oscillation code \citep[LOSC;][]{ScuflaireOsc} to compute the models and the adiabatic frequencies used in the modelling procedure both in the AIMS software \citep{Rendle2019} and in the Levenberg-Marquardt algorithm as well as in the inversion procedures discussed in Section \ref{SecStrucInv}. The minimization procedure both in AIMS and in the Levenberg-Marquardt algorithm used $4$ free parameters: the mass, the age and the initial chemical composition ($X_{0}$ and $Z_{0}$) of Kepler-444.

The forward modelling of Kepler-444 is carried out using a combination of classical and seismic data. We used the individual frequencies corrected for the line-of-sight Doppler velocity shifts \citep[see][]{Davies2014} provided in \citet{Campante2015}, which have been obtained from the observations of Kepler-444 during the nominal phase of the \textit{Kepler} mission. In addition, we used the seismic $\log g$ value from \citet{Campante2015}, the $\rm{T}_{\rm{eff}}$ value from \citet{Campante2015} as well as the value from \citet{Mack2018}. We considered the $\left[ \mathrm{Fe}/\mathrm{H} \right]$ and $\left[ \alpha / \mathrm{Fe}  \right]$ values from \citet{Mack2018} for the models using $\alpha-$enriched abundance tables. For the models using solar-scaled abundance tables, we considered the correction of \citet{Salaris1993} and used $\left[ \mathrm{M}/\mathrm{H} \right] \approx -0.37 \pm 0.09$ dex, as in \citet{Campante2015}

We carried out the modelling using the metallicity correction of \citet{Salaris1993} when using solar-scaled abundance tables, we considered both the OPAL opacity tables \citep{OPAL} and the more recent OPLIB opacity tables \citep{Colgan} to test the dependence of the modelling results in the opacity tables. We also used adequately $\alpha-$enriched abundance tables and their corresponding OPAL opacity tables \citep{OPAL}. In all cases, we included low temperature opacities of \citet{Ferguson} as well as the effects of conduction following \citet{Potekhin, Cassisi}. 

In addition to these parameters, we determined the luminosity from the parallax values given in the second \textit{Gaia} data release \citep{GAIA2016,GAIA2018}. The luminosity was computed as follows

\begin{align}
\log \left(\frac{\rm{L}}{\rm{L}_{\odot}}\right) = - 0.4 \times \left(m_{\lambda} + BC_{\lambda} - 5 \times \log d + 5 - A_{\lambda} - M_{\rm{bol},\odot}\right),
\end{align}

where $m_{\lambda}$, $BC_{\lambda}$, and $A_{\lambda}$ are the magnitude, bolometric correction, and extinction in a given band $\lambda$, $d$ is the distance, and we adopt $M_{\rm{bol},\odot} = 4.75$ for the Sun’s bolometric magnitude. We use the 2MASS K-band magnitude properties: the bolometric correction is derived using the code written by \citet{Casagrande2014,Casagrande2018a,Casagrande2018b}, while the extinction is inferred via the \citet{Green2018} dust map. As for the distance, we obtain it in two ways: by inverting the \textit{Gaia} parallax or using the distance published by \citet{Bailer2018}. Because the relative parallax error is very small for Kepler-444, the parallax inversion does not introduce any significant bias in the computation of the distance. This is why, in both cases, we obtain a similar value for the luminosity: L $\approx 0.39 \pm 0.02$ L$_{\odot}$, which is at the verge of the agreement with the value derived from the Hipparcos parallax of L $\approx 0.37 \pm 0.03$ L$_{\odot}$ reported in \citet{Campante2015}. Considering the revised spectroscopic parameters of \citet{Mack2018}, the luminosity derived from the \textit{Gaia} data is L $\approx 0.417 \pm 0.019$ L$_{\odot}$. This increase is due to the change in T$_{\rm{eff}}$ between the two studies that significantly impacts the bolometric correction, $BC_{\lambda}$. Indeed, \citet{Mack2018} provide a value of T$_{\rm{eff}}=5172\pm 75$ K, whereas \citet{Campante2015} derived a value of $5046 \pm 74$ K. Considering one or the other luminosity value with its nominal uncertainties may impact the model characteristics. Here, we make a compromise between the two studies by chosing a value of $0.40\pm0.04$ L$_{\odot}$ in our modelling, as no comment was made in \citet{Mack2018} regarding their discrepancies with \citet{Campante2015} in effective temperature.

\subsection{Global minimization technique}

\begin{figure}
\begin{center}
	\begin{minipage}{0.5\textwidth}
  	\includegraphics[width=0.478\linewidth]{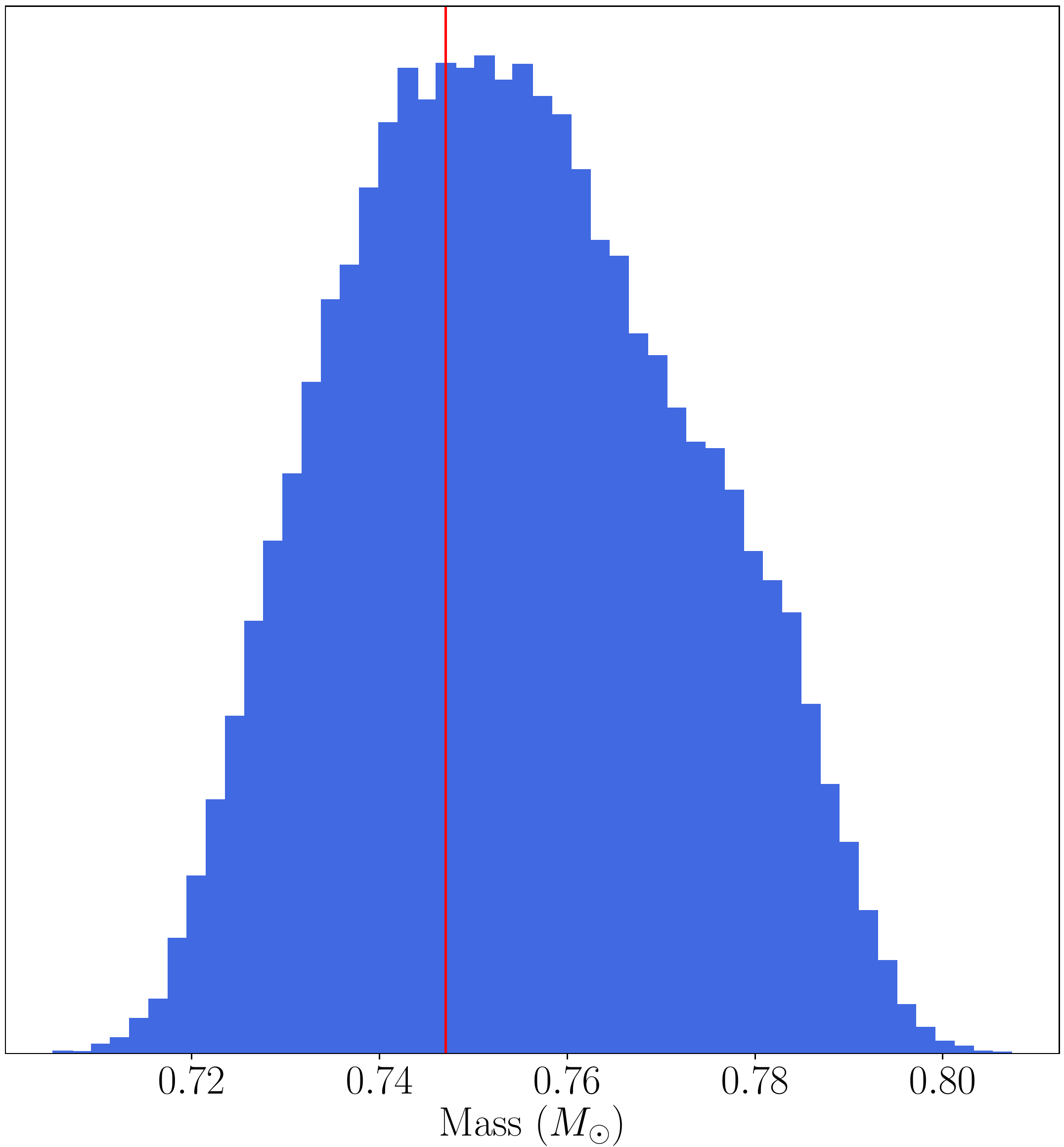} 
  	\hspace{0.02\linewidth}\includegraphics[width=0.49\linewidth]{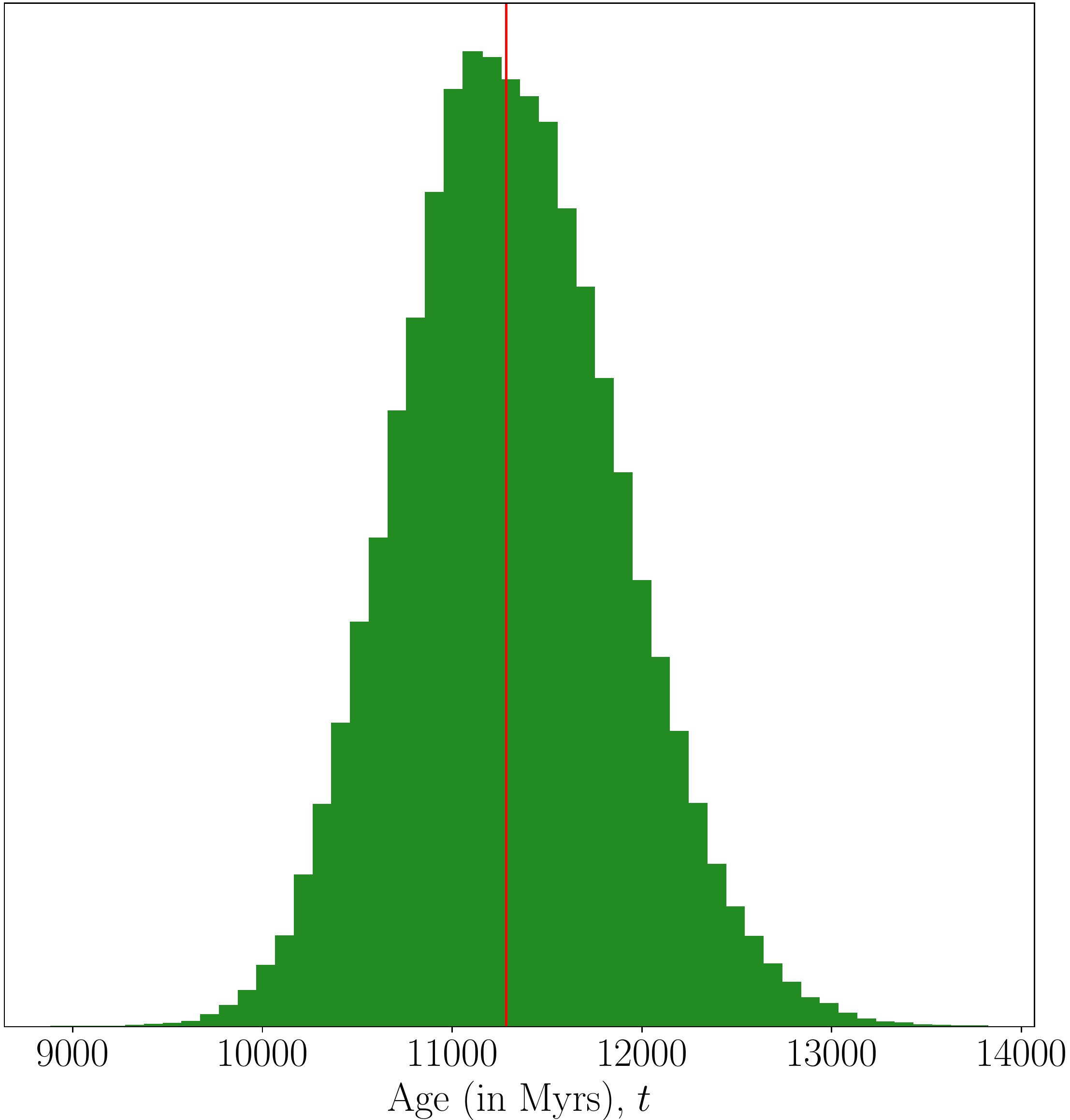}
  	\\
 	\includegraphics[width=0.48\linewidth]{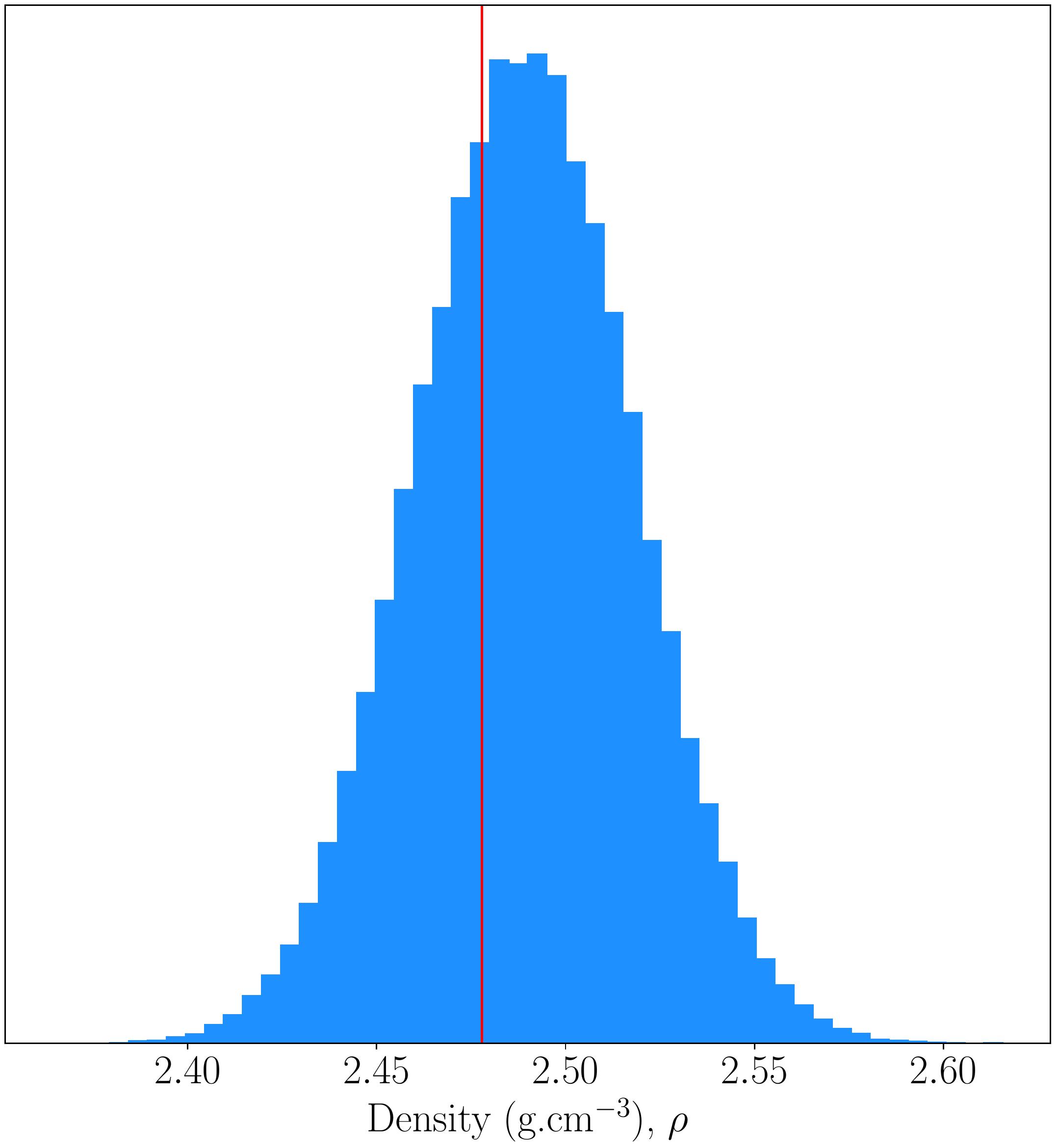}
 	\hspace{0.02\linewidth}\includegraphics[width=0.485\linewidth]{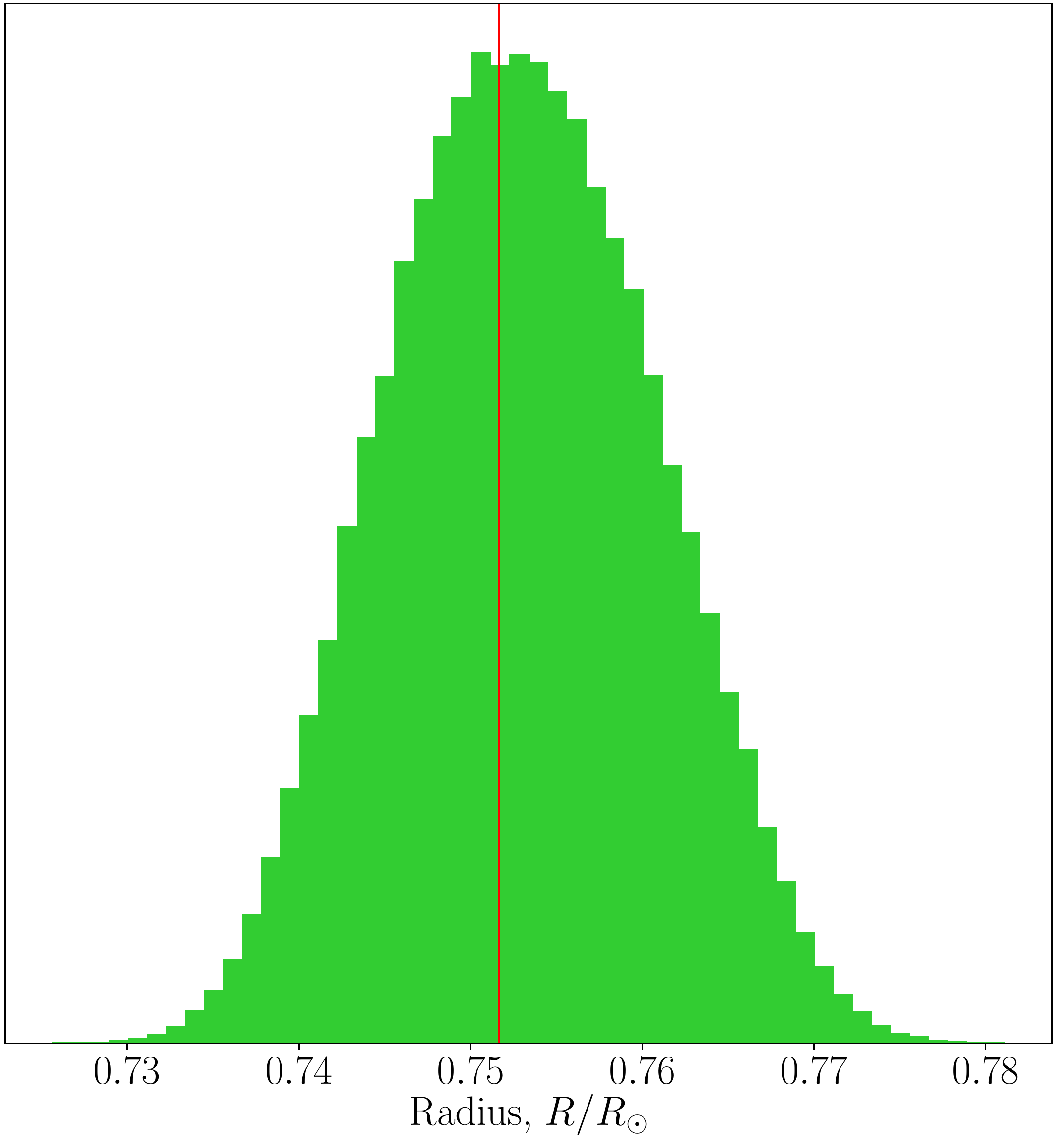}
 \end{minipage}	
 \end{center}
	\caption{Probability distribution functions for the mass (upper left), the age (upper right), the mean density (lower left), and the radius (lower right) for Kepler-444 ,obtained using AIMS. The vertical lines indicate the position of the best-model obtained from a simple scan of the grid.}
		\label{FigDistrib}
\end{figure}

First, we used the AIMS software \citep{Rendle2019} to carry out a wide-range exploration of the parameter space using different constraints for the modelling. AIMS is a global minimization tool using a Monte Carlo Markov Chain (MCMC) approach and Bayesian statistics to provide probability distribution functions for stellar parameters from a set of observational constraints. The result of one of these run is illustrated in Fig. \ref{FigDistrib} for a case where we used the following combination of seismic and classical constraints: the $r_{0,1}$ and $r_{0,2}$ frequency ratios, following the definition of \citet{RoxburghRatios}, the luminosity, the effective temperature, the seismic log g, the $\left[ \rm{Fe}/\rm{H} \right]$. We also included a mean density estimate of $2.495\pm0.050$ g/cm$^{3}$, using a reference model built with AIMS using the individual frequencies as constraints and carrying out an initial inversion as in Section \ref{SecMeanDens}. We adopted conservative error bars as no analysis of the model dependencies had been carried out at that stage. We did not rely on the results from the run using individual frequencies to determine other fundamental parameters of Kepler-444, but rather favoured the use of the frequency ratios as they are less sensitive to the surface effects. The results between the two runs were however compatible within $1.5\sigma$. The model grid, required by AIMS to carry out the minimization, was built using the FreeEOS equation of state \citep{Irwin}, the OPAL opacities, the AGSS09 abundances \citep{AGSS09}, the mixing-length theory of convection (following \citet{Cox}) and taking into account microscopic diffusion following the approach of the original work done by \citet{Thoul}, without including the \citet{Paquette} screened potentials. As upper boundary conditions, the models used an Eddington grey atmosphere \citep{Eddington1959}. 

\begin{figure}
	\centering
		\includegraphics[width=8cm]{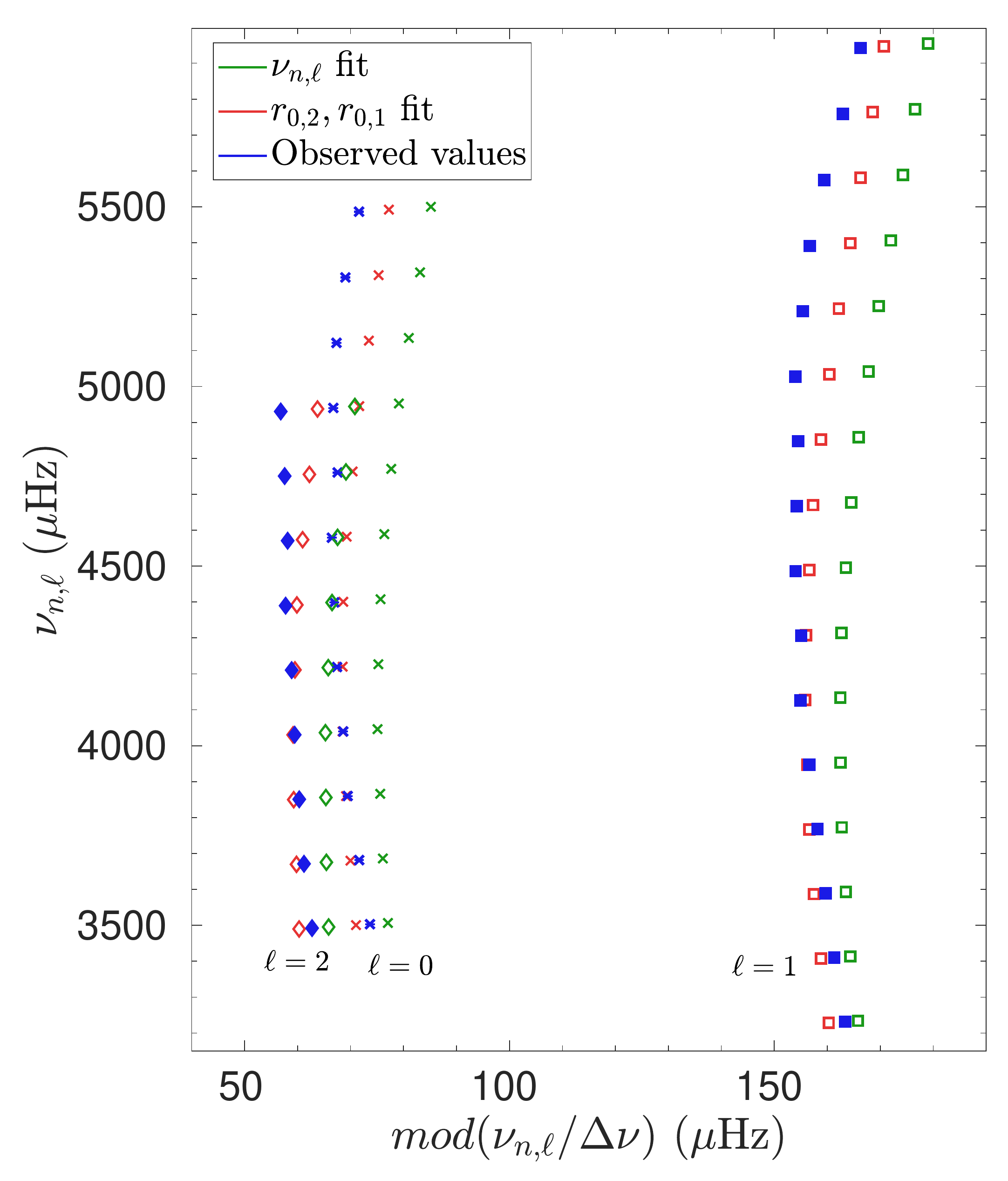}
	\caption{Echelle diagram for Kepler-444. The observational values are plotted in blue while the green and red symbols refer to two reference models computed using different seismic constraints. $\Delta \nu$ is the average large frequency separation (determined here following a least-square fitting of the individual large frequency separations for all modes).}
		\label{FigEchelle444}
\end{figure}

An illustration of the impact of using different observational constraints is shown in Fig. \ref{FigEchelle444}, where we show the results for models built using either the individual frequencies or the frequency ratios besides the non-seismic parameters listed before. As can be seen, some differences between the model frequencies are observed. The model built fitting the individual frequencies included the \citet{Ball2} correction for the surface effects. We plot here the uncorrected frequencies for this model, to illustrate the impact of the surface correction at the higher frequencies. The amplitude of the empirical surface correction is actually much larger than the observational uncertainties on the individual frequencies themselves. The model built using the frequency ratios does not show the same behaviour, but has a slightly worse agreement at low frequencies. 

However, both models have similar fundamental parameters. The main difference stems in the estimated precision of these parameters. Using the individual frequencies as direct constraints improves the precision by a factor $2.5$ compared to the one obtained from the frequency ratios. Similar effects were seen in \citet{Rendle2019} when testing AIMS on theoretical models and on the Sun. In practice, such a precision is unrealistic as the individual frequencies are strongly affected by the surface corrections. Moreover, from a theoretical point of view, each frequency does not provide an independent information on the internal structure of the star as the link between pulsation frequencies and the internal structure of a star can be expressed using only a few parameters\footnote{Typically, seismic indices such as the large and small frequency separation for pressure modes or the period spacing for gravity modes will provide the link between seismic data and stellar sructure in the asymptotic regime.}. Consequently, directly using them as constraints in a stellar model selection procedure will lead to an overestimated precision on the inferred stellar model parameters. As we will see below, changing the physical ingredients may lead to modifications of the fundamental parameters at a non-negligible order of magnitude with respect to the precision of inferences from \textit{Kepler} data. This casts somes doubts on the level of precision announced by the modelling of the individual frequencies. 

This first step of modelling using AIMS allowed us to derive a limited subspace of stellar parameters on which additional investigations can then be attempted with local minimization techniques. Hence, to fully explore the dependencies of the derived stellar parameters on the model input physics, we carried out an additional exploration of the parameter space using a Levenberg-Marquardt algorithm. This allowed us to refine the results obtained with the global minimization and compute a wide range of models on the fly, using different physical ingredients.  

\subsection{Local minimization and impact of model ingredients}\label{secLocal}

We considered a wide range of physical ingredients for the models built for the local analysis using the Levenberg-Marquardt algorithm. For the equation of state, we tested the CEFF \citep{CEFF}, FreeEOS \citep{Irwin} and OPAL \citep{Rogerseos} equations of state. For the abundances, we tested the AGSS09 \citep{AGSS09}, GN93 \citep{GrevNoels} and $\alpha$-enriched AGSS09 tables with $\left[\alpha /\rm{Fe}\right]=0.2$ and $0.3$. We considered various opacity tables, namely the OPAL \citep{OPAL} and OPLIB tables \citep{Colgan} for a scaled solar mixture, as well as OPAL tables for $\alpha$-enriched abundances. We also investigated the impact of varying the hypotheses linked to the transport of chemicals by microscopic diffusion, namely the use of the \citet{Paquette} screened potentials as well as the impact of the hypothesis of partial ionization, which ought to have an impact on the evolutionary track \citep{Schlattl}. We also varied the outer boundary conditions of the models, testing the impact of \citet{Krishna1966}, \citet{Vernazza1981} and Eddington grey $T(\tau)$ relations on the results. The convection theory of \citet{Canuto} was also used in a model and compared to the classical mixing-length theory, implemented in CLES following the approach described in \citet{Cox}. We used a solar-calibrated mixing length parameter for all our minimizations, adapting it accordingly to the outer boundary conditions and convection formalism used in the modelling. 

The fundamental parameters obtained using these various physical ingredients are summarized in Table \ref{tabModels}\footnote{MLT denotes the use of the classical mixing-length theory as in \citet{Cox}, whereas CM denotes the use of the \citet{Canuto} formalism of convection. PartIon implies the use of partial ionization in the treatment of microscopic diffusion, OV implies the inclusion of convective overshooting in the core. VAL-C and Krishna-Swamy denote the use of a \citet{Vernazza1981} and a \citet{Krishna1966} $T(\tau)$ relations.}. The minimization procedure was carried out starting from various initial conditions, inside and outside of the range provided by AIMS, for the different sets of ingredients, using the same set of classical constraints as in AIMS and the frequency ratios. Overall, the Levenberg-Marquardt minimization confirmed the values provided with the MCMC approach. 

Some minimizations including $\alpha_{\rm{MLT}}$ as one of the free parameters were also performed. We found that the value remained close to solar and did significantly improve the fit. The degeneracy of the problem did not allow us to find an ``optimal'' value of $\alpha_{\rm{MLT}}$ from the modelling. Also, following the analytical formulas of \citet{Magic} would lead to slightly increase the mixing-length parameter with respect to the solar value. This would lead the optimal model to lie within the ``higher-mass'' range of the interval we find. 

\begin{table*}[t]
\caption{Parameters of the models of Kepler-444 computed in this study.}
\label{tabModels}
  \centering
  \resizebox{\linewidth}{!}{%
\begin{tabular}{r | c | c | c | c | c | c | c | c | c | c | c }
\hline \hline
\textbf{Name}&\textbf{Mass} ($\rm{M}_{\odot}$)&\textbf{Radius} ($\rm{R}_{\odot}$) &\textbf{Age} (Gyr)&\textbf{$X_{0}$}&\textbf{$Z_{0}$}&\textbf{EOS}&\textbf{Opacity}&\textbf{Abundances} & \textbf{Diffusion} & \textbf{Convection}& \textbf{Atmosphere}\\ \hline
Model$_{1}$&$0.752\pm 0.019$&$0.748\pm0.006$&$11.57\pm0.59$&$0.747$&$0.0082$&FreeEOS&OPAL& AGSS$09$ & Thoul & MLT &Eddington\\
Model$_{2}$&$0.756\pm 0.017$&$0.754\pm0.006$&$11.39\pm0.48$&$0.733$&$0.0116$&FreeEOS&OPAL& GN$93$ & Thoul &   MLT &Eddington\\
Model$_{3}$&$0.751\pm 0.018$&$0.752\pm0.006$&$11.53\pm0.61$&$0.742$&$0.0083$&OPAL&OPAL& AGSS$09$ &   Thoul&   MLT &Eddington\\
Model$_{4}$&$0.756\pm 0.021$&$0.754\pm0.007$&$11.68\pm0.62$&$0.745$&$0.0085$&CEFF&OPAL& AGSS$09$ &  Thoul &   MLT &Eddington\\
Model$_{5}$&$0.742\pm 0.019$&$0.749\pm0.006$&$11.42\pm0.47$&$0.748$&$0.0083$&FreeEOS&OPLIB& AGSS$09$  & Thoul & MLT & Eddington \\
Model$_{6}$&$0.737\pm 0.020$&$0.752\pm0.007$&$11.72\pm0.53$&$0.742$&$0.0074$&FreeEOS&OPAL& AGSS$09$ & Thoul &  CM & Eddington\\
Model$_{7}$&$0.747\pm 0.017$&$0.749\pm0.006$&$11.20\pm0.44$&$0.750$&$0.0073$&FreeEOS&OPAL& AGSS$09$ &  Thoul &  MLT & Krishna Swamy \\
Model$_{8}$&$0.746\pm 0.019$&$0.750\pm0.006$&$11.66\pm0.60$&$0.746$&$0.0079$&FreeEOS&OPAL& AGSS$09$ &  Thoul &  MLT & VAL-C \\
Model$_{9}$&$0.747\pm 0.018$&$0.746\pm0.006$&$11.73\pm0.54$&$0.741$&$0.0084$&FreeEOS&OPAL& AGSS$09$ &  Paquette &  MLT &Eddington \\
Model$_{10}$&$0.750\pm 0.022$&$0.746\pm0.007$&$11.72\pm0.62$&$0.745$&$0.0082$&FreeEOS&OPAL& AGSS$09$ &  Paquette + PartIon &  MLT &Eddington \\
Model$_{11}$&$0.750\pm 0.019$&$0.751\pm0.006$&$11.16\pm0.55$&$0.743$&$0.0076$&FreeEOS&OPAL& AGSS$09$+$\left[ \alpha/\mathrm{Fe}\right] = 0.2$ &  Paquette + PartIon &  MLT &Eddington \\
Model$_{12}$&$0.753\pm 0.018$&$0.752\pm0.006$&$11.13\pm0.57$&$0.746$&$0.0076$&FreeEOS&OPAL& AGSS$09$+$\left[ \alpha/\mathrm{Fe}\right] = 0.3$&  Paquette + PartIon &  MLT &Eddington \\
Model$_{13}$&$0.755\pm 0.020$&$0.753\pm0.007$&$10.95\pm0.61$&$0.750$&$0.0069$&FreeEOS&OPAL& AGSS$09$+$\left[ \alpha/\mathrm{Fe}\right] = 0.2$ &  Paquette + PartIon &  MLT + OV &Eddington \\
\hline
\end{tabular}
}
\end{table*}

First, we can notice that the variations induced by the physical parameters on the Levenberg-Marquardt results remain limited, although not negligible. For most cases, they are smaller than the uncertainties given by AIMS and those reported in \citet{Campante2015}. This is not surprising, but it should be noted that some of these variations are still significant with respect to the uncertainties provided by AIMS. The largest variations are observed when using the \citet{Canuto} formalism of convection\footnote{More explicitely, we use here the expression of the convective flux by \citet{Canuto} and the expression of the scale length by \citet{Canuto91}.}, leading to a significantly lower mass. Unsurprisingly, the OPLIB opacities also significantly alter the modelling result, leading to a lower mass value. Changing the solar mixture or using $\alpha$-enhanced mixtures does not significantly alter the parameters, as already noted by \citet{Campante2015}. However, in our test cases, using $\alpha$-enhanced tables allowed to have a much better fit of non-seismic constraints while keeping the agreement with the seismic data. 

On a sidenote, given the age and low metallicity of Kepler-444, the helium abundance is naturally bound to be close to the primordial abundance. In our study, we considered the primordial helium values determined by \citet{Peimbert2016} and \citet{Fernandez2018} of respectively $Y_{P}=0.2446\pm0.0029$ and $Y_{P}=0.244\pm0.006$ as lower boundaries for our models. This value actually imposes an upper boundary on the mass of Kepler-444 and from our investigations using a solar-calibrated mixing-length, models of $0.77\rm{M}_{\odot}$ and more could only agree with seismic data if their initial helium abundance was lower than the primordial value. This higher-mass regime could potentially be reached if a significant modification was made to the radiative opacities or if one uses a non-solar calibrated mixing-length. This is for example the case if we use a mixing-length value calibrated from the $3$D simulations, following the formulas of \citet{Magic}.

The fundamental parameters found after these investigations confirm the results of \citet{Campante2015}. However, our thorough modelling has shown that changes in the physical ingredients of the stellar model could easily induce variations of the determined stellar fundamental parameters at a level significant for the reported observational uncertainties of \textit{Kepler} data. This implies that fundamental parameter determinations of \textit{Kepler} targets must consider the systematics of the model uncertainties in order to provide reliable precision estimates to other fields of astrophysics, as is attempted in table $3$ of \citet{Silva2015}. In the particular case of Kepler-444, the modelling with AIMS using the frequency ratios can provide more realistic estimates of the uncertainties on the fundamental parameters stemming from the observational uncertainties. These can be assessed from Fig. \ref{FigDistrib}. To these estimates, we can add the maximal systematic variations found from the tests using various physics to have a more robust precision\footnote{Significant revisions of the physical ingredients in the future may of course induce additional variations, we thus try to remain conservative in the $1\sigma$ error bars we provide.}. From a simple arithmetic average of the values obtained from the Levenberg-Marquardt fits, we determine the fundamental parameters of Kepler-444 to be $0.749 \pm 0.045$ $\rm{M}_{\odot}$, $0.750\pm0.015$ $\rm{R}_{\odot}$ and $11.45\pm1.0$ Gyr for the mass, radius and age, respectively. In this last step, we have increased the error bars to take into account the dispersion we saw when varying the physical ingredients in our detailed analysis with the Levenberg-Marquardt minimization algorithm. We remain conservative with respect to the error bars provide by the Levenberg-Marquardt algorithm itself, which are typically a factor $2$ lower, for all quantities (i.e. $\approx 0.02$ $\rm{M}_{\odot}$ for the mass, $ \approx 0.007$ $\rm{R}_{\odot}$ for the radius and $\approx 0.5$ Gyr for the age). Here, we have added the dispersion between the various models in Table \ref{tabModels} to these uncertainties to obtain a more robust precision for the stellar parameters determinations. We also note that these values are in good agreement with those found using AIMS. 

In Sect. \ref{SecStrucInv}, we will carry out mean density inversions and compute a second set of models using a similar strategy to what has been shown in this section to determine updated values of the fundamental parameters. These values will be given in Sect. \ref{SecPlanetary}. Moreover, we will discuss in Section \ref{SecConvCore} some additional aspects of the modelling that have been brought to light by our extended modelling procedure.

\section{Inversions for stellar structure}\label{SecStrucInv}

To further increase the thoroughness of the seismic modelling of Kepler-444, we supplemented our forward approach with seismic inversions of global quantities, so-called indicators, as in \citet{ReeseDens,Buldgen,Buldgen2015,Buldgen2018}. In the specific case of Kepler-444, we limit ourselves to inversions of the mean density. The reason of these limitations is found in the quality of the seismic data. Indeed, in the specific case of Kepler-444, the high radial order of the observed modes and the uncertainties on the frequency values do not allow us to use other quantities such as the $t_{u}$ or the $S_{Env}$ indicator s \citep[unlike the case of the study of the $16$Cyg binary system in][]{BuldgenCygA,BuldgenCygB}. We performed tests on one core condition indicator, denoted $S_{Core}$ in \citet{Buldgen2018} and found that it confirmed the properties of the models already fitting the frequency ratios. 

The inversions of structural indicators are based on the linear approximation between relative frequency differences and structural corrections derived by \citet{Dziemboswki90} and used for helioseismic inversions. The linear formulation of the inverse problem results from the variational principle of adiabatic stellar oscillations \citep{LyndenBell}. In this context, the linear relation writes
\begin{align}
\frac{\delta \nu_{i}}{\nu_{i}}=\int_{0}^{R}K^{i}_{s_{1},s_{2}}\frac{\delta s_{1}}{s_{1}}dr + \int_{0}^{R}K^{i}_{s_{2},s_{1}}\frac{\delta s_{2}}{s_{2}}dr + \mathcal{F}(\nu), \label{eqKerInversion}
\end{align}
with the $K^{i}_{s_{j},s_{k}}$ being the structural kernel functions, related to the reference model used for the inversion and the eigenfunctions of the oscillation modes. $\mathcal{F}$ is the function related to the surface effect correction. The relative differences in quantities denoted as ``$\delta$'' follow the definition
\begin{align}
\frac{\delta x}{x}=\frac{x_{obs}-x_{ref}}{x_{ref}},
\end{align}
where the subscript ``$ref$'' is related to quantities of the reference model (frequencies, sound speed or density for example) and the subscript ``$obs$'' denotes the quantities related to the observed star. The goal of the inversion procedure is to estimate the observed structural quantities $(s_{1,obs},s_{2,obs})$ from a given set of observed frequencies $(\nu_{1,obs},\nu_{2,obs})$.

In practice, Eq. \ref{eqKerInversion} can be written for a large variety of structural variables, for which corrections can then be inferred. The variables will always come in pairs, such as for example: $(c^{2},\rho)$, $(c^{2},\Gamma_{1})$, $(\rho,Y)$, $(u,Y)$ or $(S_{5/3}, Y)$ with $\Gamma_{1}=\left( \frac{\partial \ln P}{\partial \ln \rho} \right)_{S}$, $c^{2}=\frac{\Gamma_{1}P}{\rho}$ the squared adiabatic sound speed, with P the local pressure and $\rho$ the local density, $Y$ the helium mass fraction, $u=\frac{P}{\rho}$ and $S_{5/3}=\frac{P}{\rho^{5/3}}$, a proxy of the entropy of the stellar material. In those last three cases, one assumes the equation of state of the stellar material to be known and thus the relative differences in $\Gamma_{1}$ can be developed linearly according to
\begin{align}
\frac{\delta \Gamma_{1}}{\Gamma_{1}}=&\left(\frac{\partial \ln \Gamma_{1}}{\partial \ln P}\right)_{\rho,Y,Z}\frac{\delta P}{P} + \left(\frac{\partial \ln \Gamma_{1}}{\partial \ln \rho}\right)_{P,Y,Z}\frac{\delta \rho}{\rho} \nonumber \\ & + \left(\frac{\partial \ln \Gamma_{1}}{\partial Y}\right)_{P,\rho,Z}\delta Y + \left( \frac{\partial \ln \Gamma_{1}}{\partial Z} \right)_{P,\rho,Y} \delta Z,
\end{align}
with $Z$ the heavy elements abundance. In practice, this last term is often omitted in asteroseismic inversions due to the low amplitude of the metallicity differences compared to the other terms. In general, the use of $\Gamma_{1}$ or $Y$ as a secondary variable can help mitigate the amplitude of the cross-term of the inversion and thus increase the accuracy of the inference \citep[see][for a discussion in the case of asteroseismic inversions]{Buldgen2015}. 

The surface correction term, $\mathcal{F}$, is considered to be a slowly varying function of the frequency and modelled using empirical formulations \citep[see e.g.][]{Ball1,Sonoi,Ball2}. In the following section, we will test various approaches to quantify the robustness of our results with respect to the surface corrections. 

In this study, we use the Substractive Optimally Localized Averages (SOLA) inversion technique \citep{Pijpers} to determine structural indicators for Kepler-444. Other approaches have also been used to solve the inverse problem, complementary to that using the variational formulation \citep[see e.g.][]{Roxburgh2002a,Roxburgh2002b,Roxburgh2002c,Appourchaux2015}. These could potentially provide a strong complement to the linear formalism, which is intrinsically limited and can be model-dependent. 

\subsection{Mean density inversions}\label{SecMeanDens}

Mean density inversions have been formalized by \citet{ReeseDens} as a way to exploit individual oscillation frequencies to extract this key quantity from seismic observations in a model-independent way. The method has since been thoroughly tested \citep{ReeseDens,Buldgen,Buldgen2019} and applied to various cases \citep[see][]{BuldgenCygA,BuldgenCygB,Buldgen2017}. The inversion procedure is based on Eq. \ref{eqKerInversion}, written for the $(\rho,\Gamma_{1})$ structural pair. 

The inversion for the mean density using the SOLA technique is carried out by defining a target function related to the integral definition of the mean density
\begin{align}
\frac{\delta \bar{\rho}}{\bar{\rho}}=\frac{3}{4\pi R^{3}\bar{\rho}}\int_{0}^{R}4\pi r^{2}\delta \rho dr. \label{eq:EqBasisMeanDens}
\end{align}
From Eq. \ref{eq:EqBasisMeanDens}, non-dimensionalizing the integral and slightly modifying the expression gives the definition of the target function of the inversion
\begin{align}
\mathcal{T}_{\bar{\rho}}=4\pi x^{2} \frac{\rho}{\rho_{R}},
\end{align}
with $x=r/R$ the normalized radial position of an element of stellar matter, $R$ the photospheric radius and $\rho_{R}=M/R^{3}$, with $M$ the mass of the star. 

Following these definitions, the cost function of the SOLA inversion then reads
\begin{align}
\mathcal{J}_{\bar{\rho}}(c_{i})=&\int_{0}^{1}\left[K_{\mathrm{Avg}} - \mathcal{T}_{\bar{\rho}}\right]^{2}dx + \beta \int_{0}^{1}\left( K_{\mathrm{Cross}}\right)^{2}dx \nonumber \\ &+ \lambda \left[ 2 -\sum_{i}c_{i} \right] + \tan \theta \frac{\sum_{i}\left(c_{i}\sigma_{i}\right)^{2}}{\langle\sigma^{2}\rangle} + \mathcal{F}_{\mathrm{Surf}}(\nu). \label{eq:CostSOLA}
\end{align}
In Eq. \ref{eq:CostSOLA}, we have defined the averaging and cross-term kernels of the inversion, which are related to the structural kernels
\begin{align}
K_{\mathrm{Avg}}=\sum_{i}c_{i}K^{i}_{\rho,\Gamma_{1}}, \\
K_{\mathrm{Cross}}=\sum_{i}c_{i}K^{i}_{\Gamma_{1},\rho}.
\end{align}
We have also introduced the so-called trade-off parameters, $\beta$ and $\theta$ which are used to adjust the balance between fitting the target function, mitigating the amplitude of the cross term contribution of the inversion and the amplification of observational error bars of the individual frequencies ($\sigma_{i}$). In addition, we have defined $\langle\sigma^{2}\rangle=\frac{1}{N}\sum_{i=1}^{N}\sigma^{2}_{i}$ with $N$ the number of observed frequencies and introduced the inversion coefficients $c_{i}$ and a Lagrange multiplier $\lambda$.

The third term of Eq. \ref{eq:CostSOLA} stems from homologous reasoning \citep[see][]{ReeseDens}, which defines a crude non-linear generalization of the method to determine $\bar{\rho}_{\mathrm{Inv}}$, the inverted mean density, from
\begin{align}
\bar{\rho}_{\mathrm{Inv}}=\left(1+\frac{1}{2}\sum_{i}c_{i}\frac{\delta \nu_{i}}{\nu_{i}}\right)^{2}\bar{\rho}_{\mathrm{Ref}},
\end{align}
with $\bar{\rho}_{\mathrm{Ref}}$ the mean density of the reference model and $\frac{\delta \nu_{i}}{\nu_{i}}$ the relative differences between the observed and theoretical frequencies defined as in Eq. \ref{eqKerInversion}. In this study, we will apply the non-linear formula to all inversion procedures. Using this formalism, the error propagation of the inversion is also modified and thus the uncertainties on the inverted mean density are given by
\begin{align}
\sigma_{\bar{\rho}_{\mathrm{Inv}}}=\bar{\rho}_{\mathrm{Ref}}\left(1+\frac{1}{2}\sum_{i}c_{i}\frac{\delta \nu_{i}}{\nu_{i}} \right)\sqrt{\sum_{i}c^{2}_{i}\sigma^{2}_{i}}. \label{EqSigmaInvMeanDens}
\end{align}
The last term of Eq. \ref{eq:CostSOLA} is related to the correction of surface effects. Various procedures have been suggested in the litterature to tackle this tedious issue, which results from the intrinsic nature of solar-like oscillations. In helioseismology, the surface correction function is defined as a series of Legendre polynomials, which can go up to order $6$ or $7$ depending on the considered dataset. This approach is however not suitable for asteroseismic applications, where the number of observed individual frequencies is insufficient to simultaneously fit the target function, mitigate the cross-term and the error propragation while also reproducing the expected trend of the surface correction. 

Consequently, early mean density inversions limited the surface correction to a linear term. However, comparisons in hare-and-hounds exercises showed that this approach was not robust and could bias the inverted values \citep{ReeseHH}. In practice, more recent parametrizations of the surface effect are favoured and included in the inversion procedure. \citet{Buldgen2019} investigated the effects of these corrections on mean density inversions for red-giant stars and found that they could induce slight modifications of the results. Moreover, they also investigated the impact of non-adiabatic effects of the frequencies on the inverted results. 

In this study, we test various approaches for the correction of the surface effect. The first one is to directly introduce the formulation of \citet{Ball2} or \citet{Sonoi} in the SOLA cost-function and consider the coefficients of these correction laws to be free parameters to be determined by the inversion procedure, as is done for the classical helioseismic correction approach. Another approach is to correct a priori for the surface effects using either the empirical law depending on $T_{\rm{eff}}$ and $\log g$ described in \citet{Sonoi} or using pre-determined values of the coefficients of the \citet{Ball2} formula.  

\begin{figure}
	\centering
		\includegraphics[width=8cm]{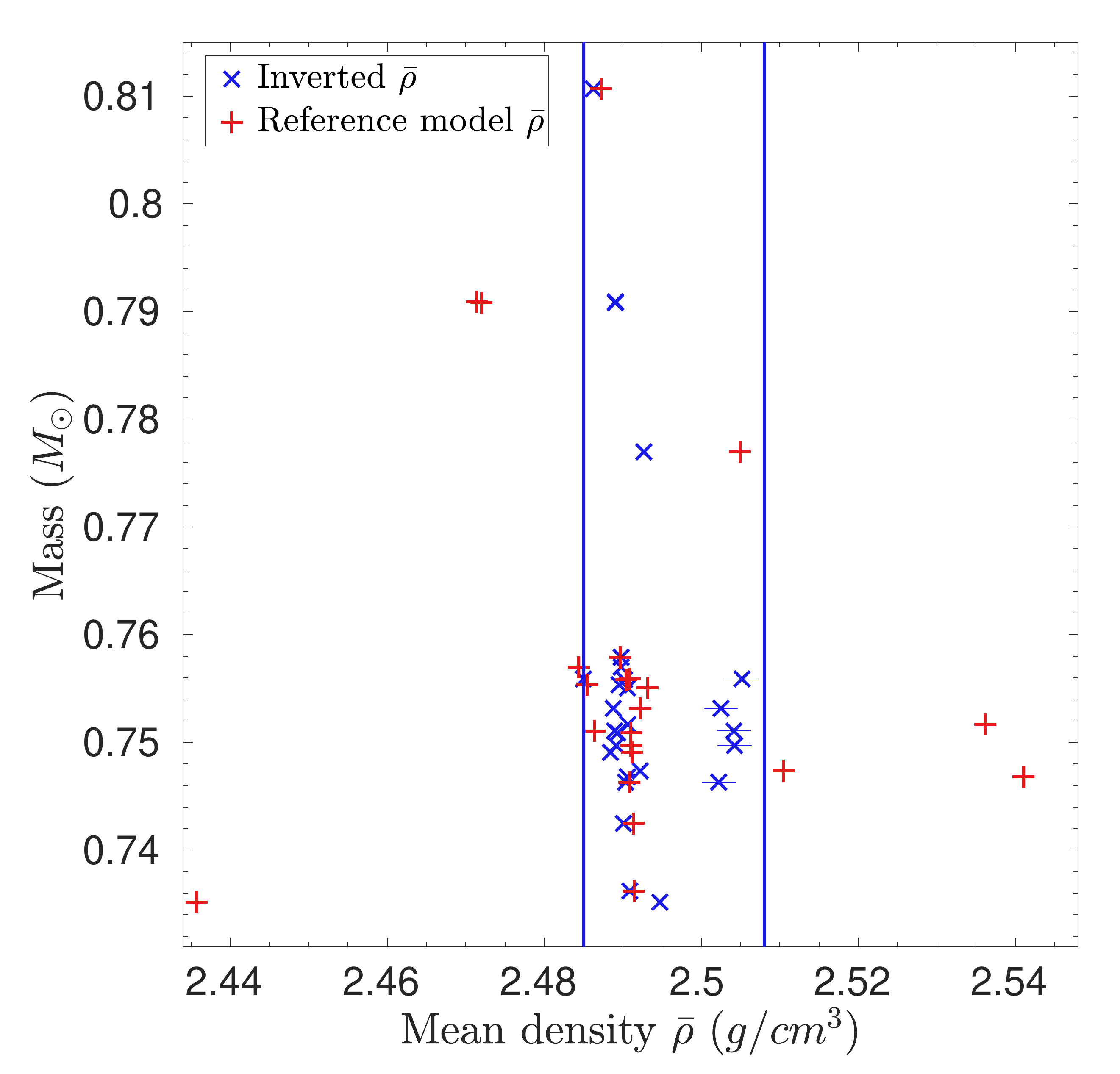}
	\caption{Results of mean density inversion for various reference models and surface corrections plotted against the masses of the reference models. The vertical lines indicate the limiting values considered to be consistent with the inversion procedure.}
		\label{FigInvRes}
\end{figure}

In the case of Kepler-444, we computed the mean density inversions for a wide range of reference models described in Sec. \ref{SecForward}, varying the correction for the surface effects. The inversion results are illustrated in Fig. \ref{FigInvRes}. As can be seen, the inversion is robust well within $1\%$, despite changing the reference model and the surface correction applied to the frequencies. Moreover, we found that the largest variations of the inversion results are observed when changing the surface correction law. More specifically, implementing directly the \citet{Ball2} surface correction as additional free parameters leads to the largest deviations, as well as a significant reduction of the quality of the averaging kernel. This confirms the results already found in \citet{Buldgen2019} for red-giant stars. 

In Fig. \ref{FigInvRes}, the effect of the \citet{Ball2} surface correction law can be seen for the inverted points with larger error bars that are above $2.5$ g/cm$^{3}$. Typically, using the \citet{Ball2} surface correction induces a shift of $0.008$ g/cm$^{3}$, whereas the \citet{Sonoi} formula induces a shift of $0.005$ g/cm$^{3}$. Both methods shift the inversion results in the same direction, namely towards higher mean density values. The spread of values obtained from the model dependency is slightly smaller, typically of the order of $0.004$ g/cm$^{3}$ and located between $2.486$ and $2.490$ g/cm$^{3}$.

\begin{figure*}
	\centering
		\includegraphics[width=16cm]{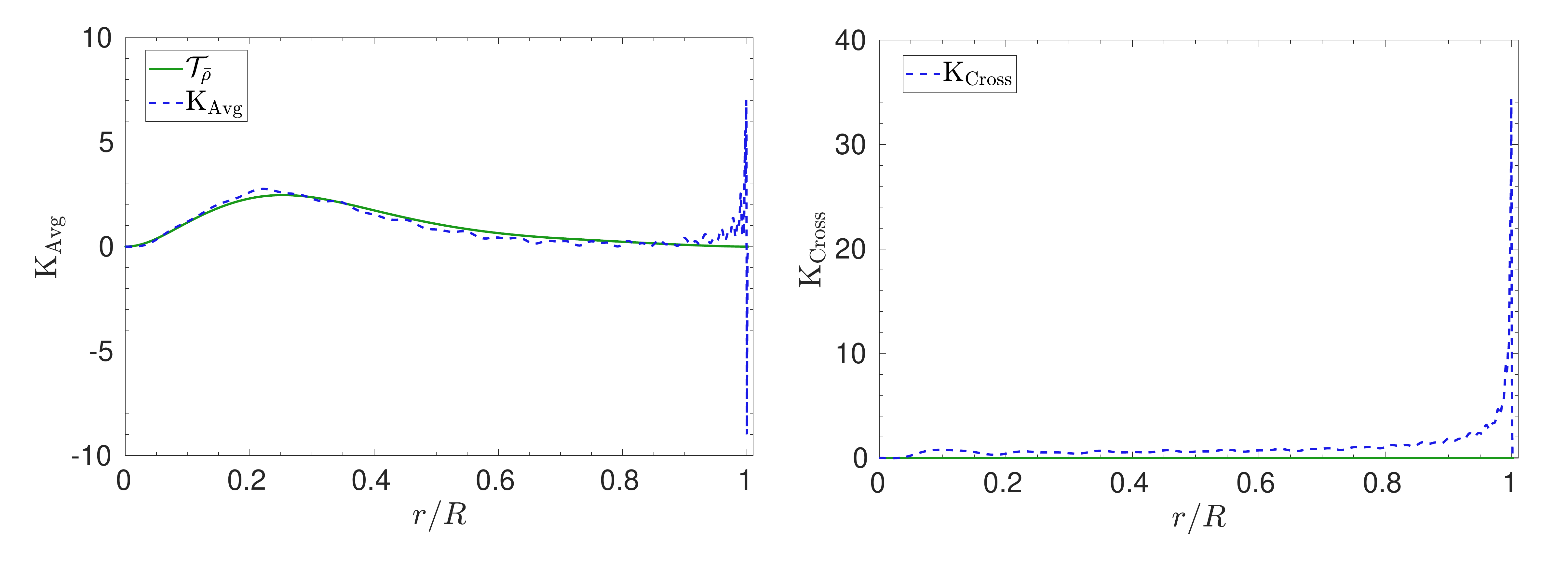}
	\caption{(Left panel) Target function of the mean density inversion (green) and averaging kernel of the inversion (blue) as a function of the adimensional stellar radius. (Right panel) Cross-term kernel of the mean density inversion (blue) and its target value, $0$, in green, as a function of the adimensional stellar radius.}
		\label{FigKernels}
\end{figure*}

Another way of assessing the robustness of the inversion is to inspect the behaviour of the averaging and cross-term kernels. This is done in Fig. \ref{FigKernels} for Model$_{1}$ of Table \ref{tabModels}. The agreement with the target function and the amplitude of the cross-term is very similar to what has been obtained in previous studies analysing the accuracy and robustness of mean density inversions\footnote{The value of the cross-term is multiplied by the relative differences in $\Gamma_{1}$ between the reference model and the target, leading typically to negligible errors in the procedure \citep[see][for a discussion]{ReeseDens,Buldgen}.}. Looking at Fig. \ref{FigInvRes}, we can state that, by using various reference models and surface effect corrections, the mean density inversion has still provided a very accurate and robust determination of the mean density of Kepler-444, beyond what would have been achievable by standard forward modelling.

Overall, the behaviour of the inversion is satisfactory. We thus conservatively assess that the mean density of Kepler-444 must lie within $2.496 \pm 0.012$ g/cm$^{3}$. This result agrees with the value reported by \citet{Campante2015}, who found $2.493\pm 0.028$ g/cm$^{3}$. Unsurprisingly, the inversion technique, making full use of the information of the frequency spectrum to determine the value of the indicator, provides a more precise estimate of the sought integrated quantity, here by a factor $2$. This value can then be used to constrain stellar and planetary parameters to a higher degree of precision. We note that some of the models computed in section \ref{SecForward} already fitted the mean density within its uncertainties. 

On a sidenote, it is also worth noticing that even the mean density inversions show some degree of model-dependency and to an even higher degree, a dependency in the surface correction approach used when computing the inversion. This confirms that asteroseismic inversions, at least within the linear formalism\footnote{Although similar issues have been reported in \citet{Appourchaux2015} for non-linear inversion.}, are not fully model-independent. Consequently, assessing their true precision cannot be done by merely propagating the observational error bars of the individual frequencies derived from Eq. \ref{EqSigmaInvMeanDens}.

\subsection{Revised forward modelling and impact on planetary parameters}\label{SecPlanetary}

Thanks to the very precise determination of the mean density of Kepler-444 from the inversion procedure, we can re-run a set of models using the Levenberg-Marquardt minimization and determine updated values for the fundamental parameters. We used the same set of constraints as in Section \ref{SecForward}, but reduced the uncertainties on the mean density from $0.05$ to $0.012$ g/cm$^{3}$, in agreement with the inversion procedure. As stated in Section \ref{SecMeanDens}, some of the reference models already agreed with the inverted mean density within the precision of the inversion procedure. However, this was not the case for all models, particulary those responsible for the largest spread in fundamental parameters such as those using the \citet{Canuto} formalism of convection or the OPLIB opacity tables. Whenever possible, we used the $\alpha$-enriched opacities and abundance tables to compute this second set of models.

While the mean values of the fundamental parameters did not change significantly, recomputing the whole set of models using this more accurate value of the mean density allowed for a higher precision of their determination. The final values for this consolidated stellar seismic modelling are summarized in table \ref{tabFinalStar}. For this last set of values, we have still kept the error bars conservative and combined the errors from the remaining dispersion between the various models and those from the uncertainties on the observational constraints. We note that these final values are in excellent agreement with those found in \citet{Campante2015}. They are however, slightly more precise, as \cite{Campante2015} report a $0.043$ $M_{\odot}$ uncertainty on mass, a $0.013$ $R_{\odot}$ uncertainty on radius and a $+0.91$ and $-0.99$ Gy uncertainty on age. This is likely a consequence on the improved precision on the stellar mean density as a result of the use of the seismic inversion as we used a slightly larger uncertainty on the luminosity than \cite{Campante2015}, as a result of the disagreements in $T_{\rm{eff}}$ between the recent determinations found in the litterature. 

\begin{table}[t]
\caption{Improved stellar parameter values for Kepler-444.}
\label{tabFinalStar}
  \centering
\begin{tabular}{r | c | c }
\hline \hline
\textbf{Mass}($\rm{M}_{\odot}$)&\textbf{Radius} ($\rm{R}_{\odot}$) &\textbf{Age} (Gy) \\ \hline
$0.754 \pm 0.030$ $\rm{M}_{\odot}$&$0.753 \pm 0.010$ $\rm{R}_{\odot}$&$11.00 \pm 0.8$\\
\hline
\end{tabular}
\end{table}

\begin{figure}
	\centering
		\includegraphics[width=8cm]{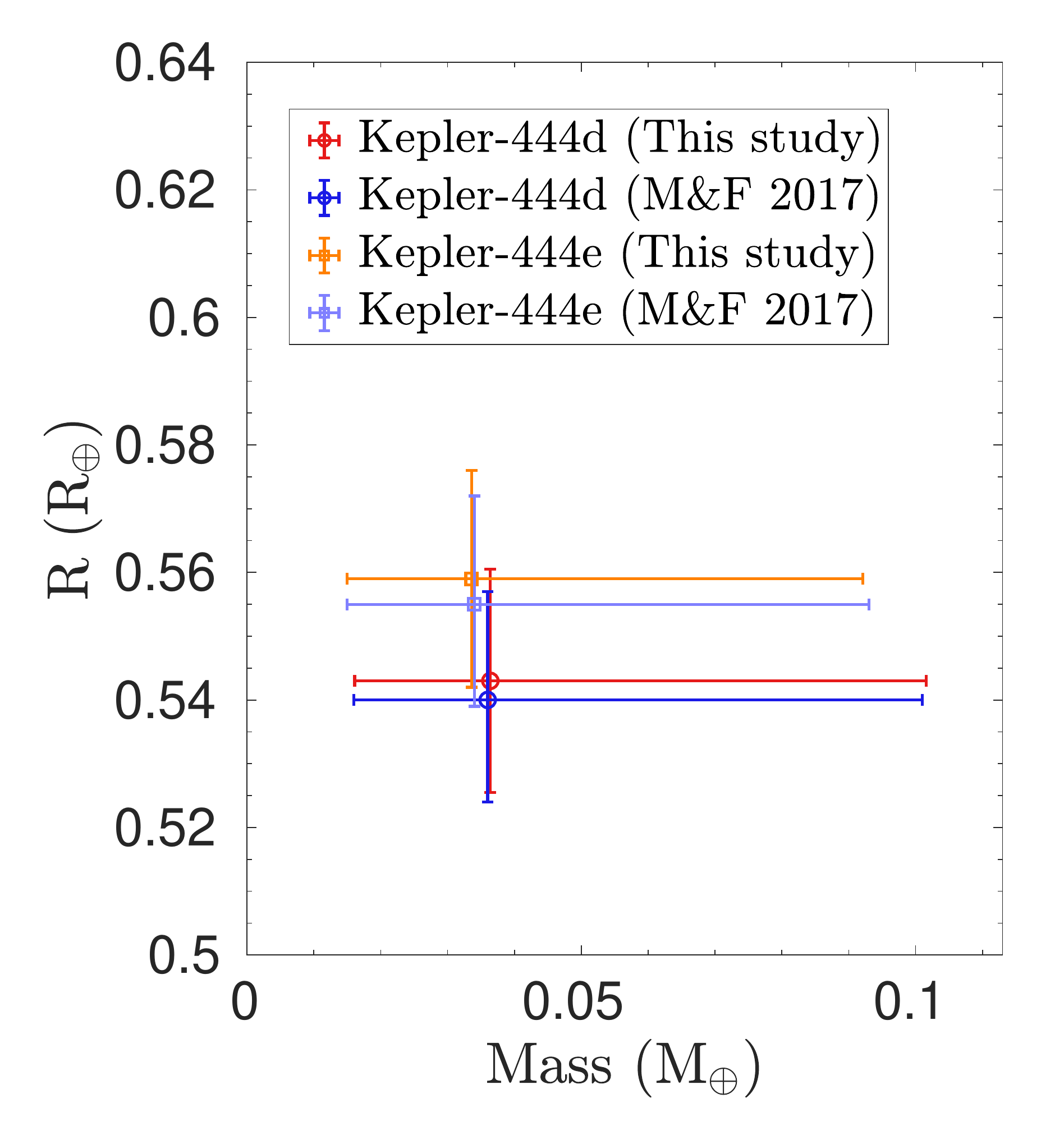}
	\caption{Comparisons between the masses and radii of Kepler-444d and Kepler-444e  determined from this study and the study of \citet{Mills2017}.}
		\label{FigMR}
\end{figure}

Using these updated stellar parameters, we revise the most recent estimates of the planetary parameters derived by \citet{Mills2017}. These results are summarized in Table \ref{tabPlanet} and we illustrate the position of planet e and f on a mass-radius diagram in Fig. \ref{FigMR}. 

\begin{table}[t]
\caption{Revised planetary parameters for Kepler-444 system.}
\label{tabPlanet}
  \centering
\begin{tabular}{r | c | c }
\hline \hline
\textbf{Name}&\textbf{Mass} ($\rm{M}_{\oplus}$)&\textbf{Radius} ($\rm{R}_{\oplus}$) \\ \hline
Kepler-444b&-&$0.408\pm0.014$\\
Kepler-444c&-&$0.528\pm0.017$\\
Kepler-444d&$0.0364^{+0.0652}_{-0.0203}$&$0.543\pm0.018$\\
Kepler-444e&$0.0336^{+0.0585}_{-0.0186}$&$0.559\pm0.017$\\
Kepler-444f&-&$0.772\pm0.019$\\
\hline
\end{tabular}
\end{table}

As could already be guessed from comparisons with \citet{Campante2015}, we find no significant modifications to the values of the planetary masses and radii. Changes in the precision of the planetary radii are minimal, due to the reduced changes on the precision on the stellar radii. Similarly, the uncertainties on the masses values from \citet{Mills2017} are largely dominated by the uncertainties on the TTV data. Thus, revising the stellar parameters has little to no impact on the planetary masses. However, other systems where the uncertainties on the host star dominate could significantly benefit from a detailed seismic study as the one carried out here. Consequently, we find that Kepler-444e and Kepler-444f still remain excellent candidates of ocean planets. As noted by \citet{Mills2017}, future observations with PLATO may provide the additional data required to determine more precisely the masses and radii of the planets orbiting Kepler-444, confirming their nature.

\section{Survival of a convective core in Kepler-444}\label{SecConvCore}

\begin{figure*}
	\centering
		\includegraphics[width=16cm]{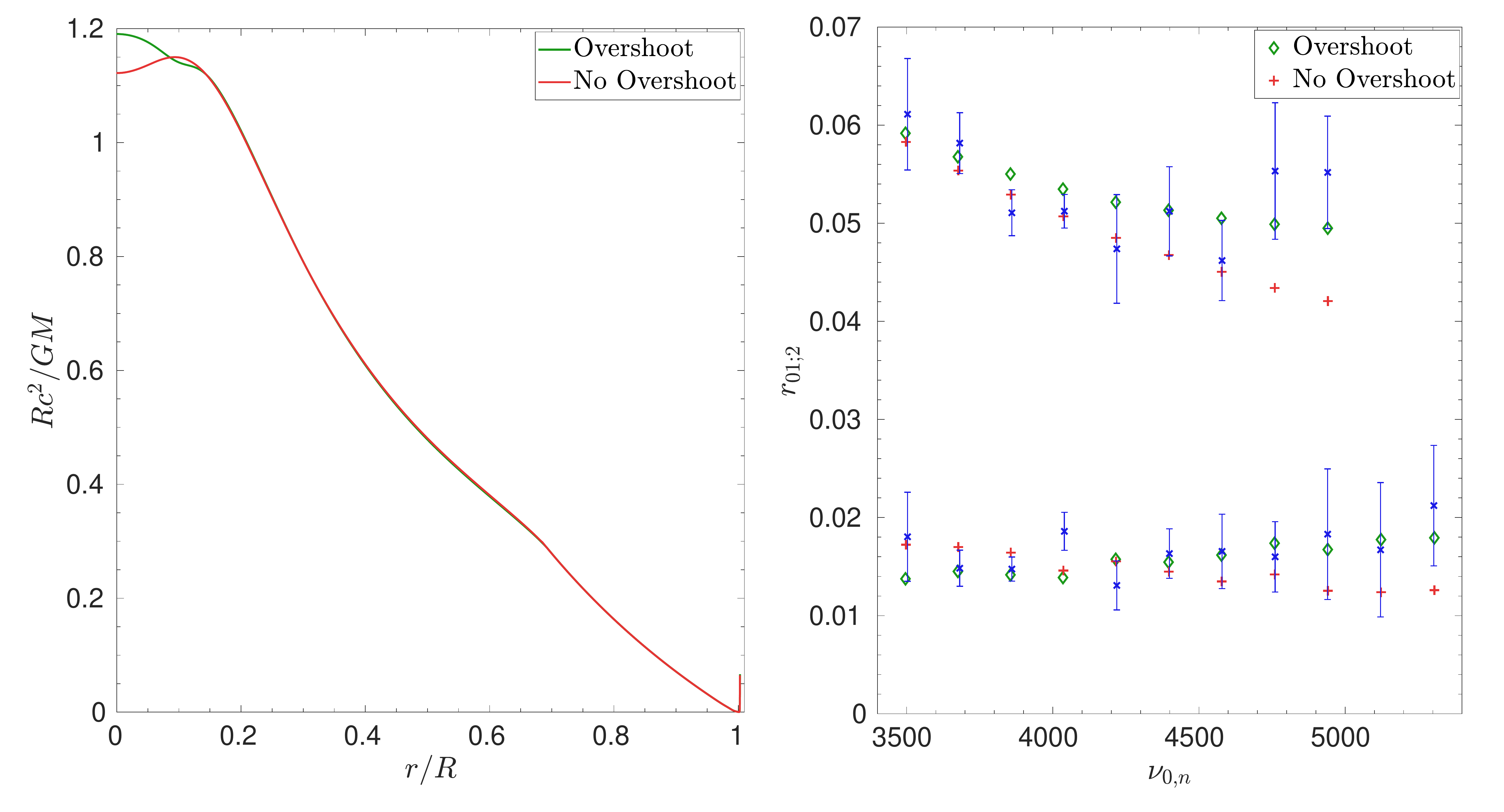}
	\caption{(Left panel) Adimensional sound speed profile as a function of the adimensional stellar radius for two models (Model$_{12}$ and Model$_{13}$ in table \ref{tabModels}) of Kepler-444 with (green) and without (red) overshooting. (Right panel) Frequency ratios as a function of the observed radial frequency. The upper symbols are related to the $r_{0,2}$ ratios while the lower ones are related to the $r_{0,1}$ ratios. The blue symbols are related to the observed values of the ratios while their theoretical counterparts are plotted in green and red for Model$_{13}$ and Model$_{12}$ with and without overshooting, respectively.}
		\label{FigRatios}
\end{figure*}

In section \ref{SecForward}, we have shown that the spread of the fundamental parameters we obtained remained small, but not negligible, compared to the uncertainties obtained from AIMS and the Levenberg-Marquardt minimization using one set of physical ingredients. The extended testing procedure we used allowed us to conclude that the derived parameters are robust. However, further investigations have led us to consider whether the accuracy of the fit could be further improved. It appeared that, despite converging systematically on the same results, all models presented a slight disagreement in the frequency ratios. We illustrate this effect in the right panel of Fig. \ref{FigRatios}, where we show in red the results for a best fit model ``Model$_{11}$'' of Table \ref{tabModels}. One can clearly see that the higher range of the frequency ratios are not well fitted by this model. This disagreement is present in all models considered, regardless of the physical ingredients used to fit the ratios.

One solution to improve the agreement with these seismic indices is to include a moderate amount of overshooting in the modelling (typically $\alpha_{Ov}\approx 0.18 \rm{H}_{P}$, with $\rm{H}_{P}$ the pressure scale height defined as $\frac{dr}{d \ln P}$). The results for such a model of Kepler-444, including overshooting, namely Model$_{13}$ of Table \ref{tabModels}, are illustrated in green in the right panel of Fig. \ref{FigRatios}. The variation in $\chi^{2}$ is relatively small, as one goes from $1.24$ for the reduced $\chi^{2}$ of a model without overshooting to $0.99$ for a model including overshooting. The largest contributions to the $\chi^{2}$ in the case with overshooting stems from the $r_{0,1}$ ratio of the $n=21$ and $n=20$ modes for the seismic constraints and from $\rm{T}_{\rm{eff}}$ for the non-seismic constraints. We note however that the values of $\rm{T}_{\rm{eff}}$ tend to favour the revised determination of \citep{Mack2018}, but no firm conclusions can be drawn given the relatively large uncertainties.

\begin{figure*}
	\centering
		\includegraphics[width=16cm]{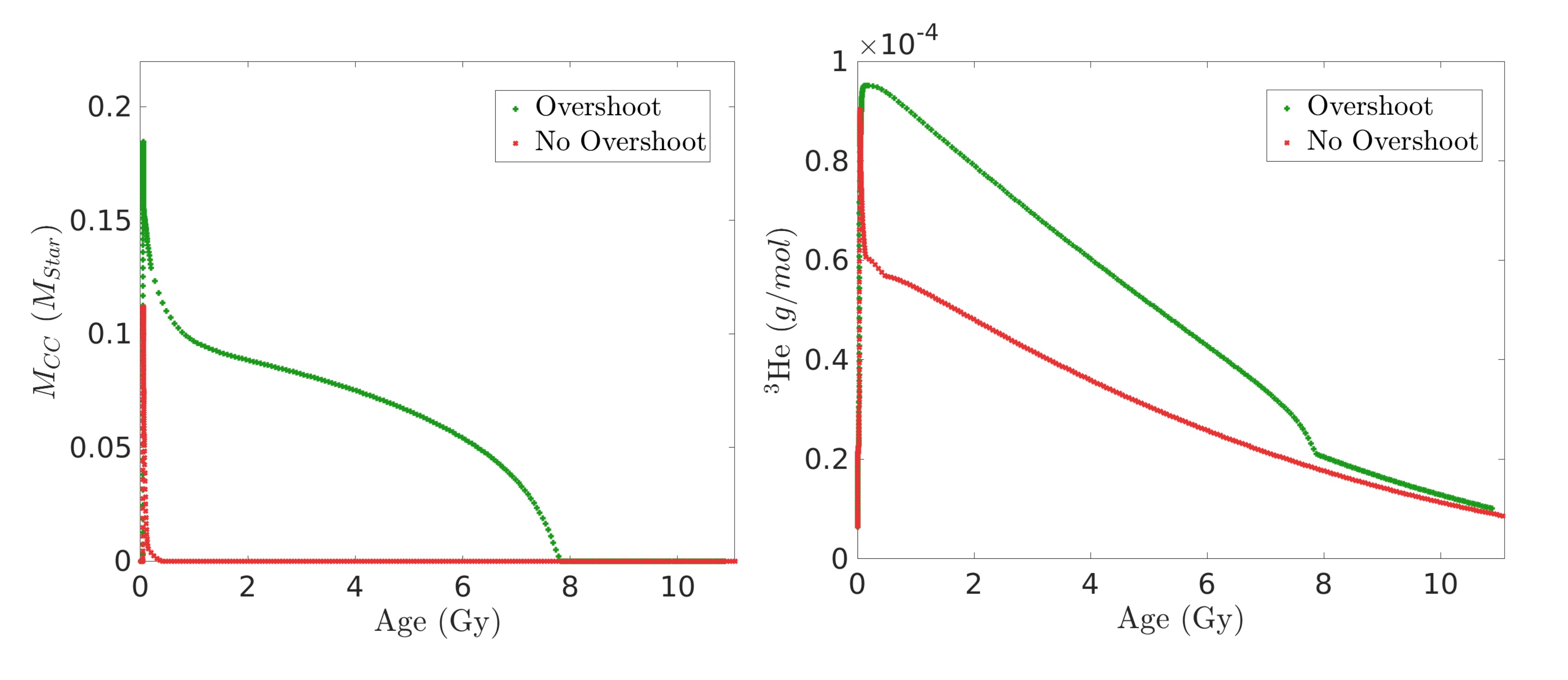}
	\caption{(Left panel) Mass of the convective core expressed as a fraction of the stellar mass in Model$_{13}$ (with overshooting, green) and Model$_{12}$ (without overshooting, red) of Kepler-444 as a function of age. (Right panel) Central $^{3}\rm{He}$ abundance as a function of age for Model$_{12}$ (red) and Model$_{13}$ (green) of Kepler-444.}
		\label{FigMMCHe3}
\end{figure*}

Further increasing the extent of the mixed region leads, in this case, to the survival of the convective core on the main sequence for the model of Kepler-444 and significantly alters its present-day sound speed profile. In the left panel of Fig. \ref{FigRatios}, we illustrate the sound speed profile of two optimal models for Kepler-444 with and without a surviving convective core on the main sequence. In the left panel of Fig. \ref{FigMMCHe3}, we illustrate the extent of the convective core as a function of the age of the model. As can be seen, including overshooting helps the convective core to maintain itself for a significant portion of the main-sequence, despite the very low mass of the star. 

Such a situation, although unexpected at a first glance, is not uncommon, as \citet{Deheuvels2010} reported a similar feature in HD$203608$, a low-mass low-metallicity, and alpha-enriched solar-like oscillator observed by CoRoT. In the case of Kepler-444, the signature of the convective core appears fainter. To further confirm this result, we computed the so-called small frequency spacing used by \citet{Deheuvels2010} to discriminate models without overshooting and tested whether these indices could also keep trace of the survival of a convective core in Kepler-444. We found that the small spacings were not as constraining as the frequency ratios. Their behaviour could be altered by other ingredients such as the upper boundary conditions and the use of the \citet{Canuto} formalism of convection in the envelope. The fact that we were not able to see the trace of the convective core in the small spacings might stem from the fact that it has already disappeared in Kepler-444 at the time of observation. This was not the case in HD$203608$, which still bears a convective core according to the seismic analysis of \citet{Deheuvels2010}. 

This behaviour can also be understood by looking at the left panel of Fig. \ref{FigRatios}, illustrating the differences in sound speed for a model with and without a long-lasting convective core of Kepler-444. By comparing this illustration to figure $2$ of \citet{Deheuvels2010}, we can see that the sharp signature of the convective core that is present in HD$203608$ is almost erased in Kepler-444, due to the effects of microscopic diffusion. Following the second order asymptotic developments of \citet{Provost1993}, \citet{Deheuvels2010} explained that the small spacing would be sensitive to sharp variations in the sound speed profile. Since this variation has disappeared in Kepler-444, it is not surprising that the small spacings are very similar between models with and without a surviving convective core on the main sequence. The slope of the sound speed profile in the core is still, however, very different between the two models, as can be seen from the left panel of Fig. \ref{FigRatios}. This explains why the frequency ratios, indicative of the sound speed gradient in deep layers, provide a clearer diagnostic of the core history of Kepler-444. 

By further increasing the values of the overshooting parameter, one can make the convective core survive even longer. However, attempts with larger values, that could lead to the presence a convective core in Kepler-444 at its current age, disagree with the observed ratios. When attempting to carry out minimizations with such larger values, the disagreement with the frequency ratios led to models for which the core had disappeared at the observation time, thus agreeing with the observed frequency ratios as models with lower overshooting values, but larger discrepancies than for global parameters such as the mean density, luminosity, log g and $\left[  \rm{Fe}/\rm{H} \right]$.

The physical explanation behind the survival of the convective core in such circumstances has been extensively detailed in \citep{Deheuvels2010}. In the case of Kepler-444, tests using models built using seismic and classical constraints with and without overshooting indicate a similar origin as in the case of HD$203608$. In Fig. \ref{FigMMCHe3}, we illustrate the evolution of the convective core in a model of Kepler-444 with and without overshooting and show its link to the evolution of the abundance of $^{3}\rm{He}$ in the central regions.

The evolution and survival of the convective core in the model is indeed clearly due to the $^{3}\rm{He}$ fusion reactions. Without overshooting, the abundances quickly reach an equilibrium value for the $\left[ ^{3}\rm{He} /  \rm{H} \right]$ ratio, and the ppI chain achieves equilibrium. If overshooting is included during the evolution, the additional mixing provides more $^{3}\rm{He}$ in the central regions and extends the duration of nuclear burning out of equilibrium. Due to the much higher temperature sensitivity of out-of-equilibrium $^{3}\rm{He}$ burning, the generated energy flux cannot be evacuated by radiation alone and the convective core lingers longer during the main sequence, up to an age of $\approx 8$ Gy. However, as the extension of the convective core is smaller than the region where nuclear burning occurs the $\left[ ^{3}\rm{He}/ \rm{H} \right]$ ratio reaches its equilibrium value after a while. The convective core then quickly disappears and nuclear burning now occurs in radiative equilibrium conditions. This can be seen in Fig. \ref{FigMMCHe3}, where the slope of the central $^{3}\rm{He}$ abundance changes drastically at the time the convective core disappears. From an inspection of the energy generation rates, we can also confirm that the CNO cycle did not play any part in sustaining the convective core, the central temperatures being too low for it to have a significant role during the evolution. 

While additional mixing is essential to sustain the convective core for an extended period on the main sequence, its origin is not necessarily only linked to ``convective overshooting''\footnote{That is, the penetration of convective elements beyond the Schwarzschild border of the convective zone}. Efficient mixing may also occur if, at the beginning of the main-sequence, the star rotates almost in a solid way. In those conditions, transport of chemicals by the meridional circulation may also provide an additional source of $^{3}\rm{He}$ that will provide material for $^{3}\rm{He}$ burning and thus sustain the transitory convective core. This phenomenon is also mentioned in \citet{Roxburgh1985} to explain a potential mechanism to sustain an efficient mixing in the solar core. The phenomenon was linked to the so-called ``solar-spoon'' mechanism presented by \citet{Dilke1972} which was extensively discussed \citep[see e.g.][for a few references]{Ulrich1973,JCD1974,Unno1975,Shibahashi1975} and also investigated for Population III stars \citep{Sonoi2012}. In the latter case, the mixing results from the transport of chemicals by the non-linear gravity modes generated by a form $\epsilon$-mechanism due to the nuclear burning of $^{3}\rm{He}$. Seismic data alone is not sufficient to distinguish between one and the other case, or more likely a combination of both effects. To that end, additional contraints on the lithium and beryllium abundances may provide an indication of the efficiency of rotational mixing in the early stages of evolution and thus inform us on whether rotation would be able to provide the required mixing of chemicals. 

As seen from the seismic modelling results, the survival of the convective core does not affect the global stellar parameters. The main reason is that its extension is smaller than the region of nuclear burning. Hence, it does not modify significantly the evolutionary track of Kepler-444. However, it could potentially affect the angular momentum transport properties and lead to discrepancies in later evolutionary stages. This aspect still remains rather speculative and will be further investigated in future studies. 

\section{Conclusion}\label{SecConc}

In this paper, we have carried out a thorough modelling of Kepler-444 using updated parallax values from \textit{Gaia} DR$2$ \citep{GAIA2018} as well as updated spectroscopic parameters \citep{Mack2018}. We combined global and local forward modelling approaches to structural inversion techniques to derive robust fundamental parameters for this exoplanet-host star. To do so, we tested a wide range of physical ingredients and analyzed the spread of the fundamental parameters obtained. Amongst others, we investigated the impact of $\alpha$-enriched abundance tables and their corresponding opacities rather than solar-scaled abundances. Overall, we confirm the results of \citet{Campante2015} and also find a relatively good agreement with the results of \citet{Silva2015}. We note, however, that the luminosity derived from the \textit{Gaia} parallax is significantly higher than those obtained in these studies, especially if we consider recent spectroscopic constraints. Some of these discrepancies stem from the \textit{Gaia} parallax itself, but also from differences in the effective temperature values reported in the literature. The case of Kepler-444 is thus a good illustration where the use of the very precise \textit{Gaia} parallaxes is limited by other spectroscopic constraints.

In addition to forward modelling, we use mean density inversions to provide a very precise, nearly model-independent value for the stellar mean density of $2.496 \pm 0.012$ g/cm$^{3}$. This value was then used as a direct constraint for a second set of forward modelling, combined with the other classical and seismic constraints used in the modelling. Using this two-step approach, we determine the mass, radius and age of Kepler-444 to be $0.754 \pm 0.030$ $\rm{M}_{\odot}$, $0.753 \pm 0.010$ $\rm{R}_{\odot}$ and $11.00 \pm 0.8$ Gyr. As these parameters are very similar to those of \citet{Campante2015}, we find no significant modifications of the planetary parameters using our modelling results.

However, we have determined that Kepler-444 bore a convective core during a significant fraction of its main-sequence evolution, as was found by \citet{Deheuvels2010} for the CoRoT target HD$203608$. The origin of the convective core in the case of Kepler-444 is similar to the case of HD$203608$ and can be linked to the nuclear burning of $\rm{He}_{3}$ out of equilibrium. The absence of equilibrium conditions for a prolongated portion of time after the ZAMS is due to an additional mixing, providing more nuclear fuel for the $\rm{He}_{3}$ combustion. The origin of this mixing is unclear and can be due to rotation, overshooting or even non-linear gravity modes, although this latter case seems less plausible from the stability analyses of \citet{Sonoi2012}. The seismic data alone cannot disentangle between these processes. Promising indicators of the efficiency of rotational mixing in the early stages that could help lift this degeneracy are the lithium and beryllium abundances, yet to be determined in the case Kepler-444. So far, using seismology, we are however able to estimate the lifetime of the convective core to be around $8$ Gyr. 

The survival of the convective core, as noted by \citet{Deheuvels2010}, does not affect the fundamental parameters of the star, but it could play a key role in the angular momentum transport in later stages of evolution. A thorough modelling of the star including rotation and transport by magnetic instabilities as in \citet{Eggenberger2010} will be undertaken in a future paper. These transport prescriptions will then be used to characterize the dynamical properties of the system, following the approach of \citet{Privitera2016I} and \citet{Rao2018}. In turn, such detailed studies will pave the way for joint analyses of the stellar-planetary system as a whole, including atmospheric characterizations and the change of these properties when coupled to seismically calibrated non-standard stellar evolution models.  

With the advent of TESS and PLATO, such synergies between exoplanetology and asteroseismology will be further reinforced. For the best targets, a thorough modelling and investigation of the history of the system with respect to the evolution of its host star can be foreseen. In this series of papers, we will illustrate these possibilities for a promising target observed by \textit{Kepler}. In the specific case of Kepler-444, \citep{Mills2017} estimate that PLATO observations could help to significantly reduce the current uncertainties on the planetary masses.

\section*{Acknowledgements}

This work is sponsored by the Swiss National Science Foundation (project number $200020-172505$). M.F. is supported by the FNRS (``Fonds National de la Recherche Scientifique'') through a FRIA (``Fonds pour la Formation à la Recherche dans l'Industrie et l'Agriculture'') doctoral fellowship.  S.J.A.J.S. is funded by the Wallonia-Brussels Federation ARC grant for Concerted Research Actions. A.M. and J.M. acknowledge support from the ERC Consolidator Grant funding scheme ({\em project ASTEROCHRONOMETRY}, G.A. n. 772293). We gratefully acknowledge the support of the UK Science and Technology Facilities Council (STFC). V.B. acknowledges by the Swiss National Science Fundation (SNSF) in the frame of the National Centre for Competences in Research PlanetS and has received fundings from the European Research Council (ERC) under the European Union's Horizon $2020$ research and innovation programme (project Four Aces; grant agreement number 724427). This article used an adapted version of InversionKit, a software developed within the HELAS and SPACEINN networks, funded by the European Commissions's Sixth and Seventh Framework Programmes.

\bibliography{biblioarticle444}

\end{document}